\documentstyle[prb,aps,twocolumn,floats]{revtex}
\def\prb{Phys. Rev. B }
\def\prl{Phys. Rev. Lett. }

\begin{document}
\input epsf
\draft
\wideabs{
\title{Thermal conductivity of YBa$_2$(Cu$_{1-x}$Zn$_x$)$_3$O$_{7-y}$ single
crystals}
\author{A.~V.~Inyushkin, A.~N.~Taldenkov, S.~Yu.~Shabanov}
\address{Institute of Molecular Physics, Russian Research Centre ``Kurchatov
Institute'',\\
123182 Moscow, Russia}
\author{L.~N.~Dem'yanets, T.~G.~Uvarova}
\address{Shubnikov Institute of Crystallography, Russian Academy of Science,\\
117333 Moscow, Russia\\
 }
\date{Superconductivity: physics, chemistry, technique {\bf 6}, 
985-997 (1993) \protect \cite{spct}\\
Received November 26, 1992; accepted April 26, 1993}
\maketitle

\begin{abstract}
The in-plane thermal conductivity $K(T)$ of YBa$_2$(Cu$_{1-x}$Zn$_x$)$_3$O$%
_{7-y}$ single crystals (YBCO:Zn) has been investigated in the temperature
range from 5 to 250~K and in magnetic fields up to 30~kOe. The thermal
conductivity decreases substantially with increase of the Zn concentration
providing reduction in the thermal conductivity peak and its disappearance
at a high content of the dopant. At temperatures below 15~K the $K(T)$
dependencies are functionally similar for all the samples. The magnetic
field effect on the thermal conductivity is strongly suppressed by doping
even in the crystal with the lowest Zn content. For YBCO:Zn with $x=0.003$ 
at temperatures below 25~K the magnitude of remnant magneto-thermal 
resistivity (after switching off the 30~kOe magnetic field) is about one 
tenth of that for pure YBCO. Analysing experimental data we use the theory 
of thermal conductivity in high-$T_{c}$ superconductors and speculations 
about the modification of quasiparticle excitation spectrum due to the 
Cooper pair breaking.
\end{abstract}

}
\narrowtext

\section{Introduction}

The results of many experimental works, e.g., tunnel \cite{Gur} and Raman 
\cite{Coop,Thom,Boek} spectroscopy studies indicate that there is nonzero 
electron density of states inside the energy gap in YBCO in the 
superconducting state. The appearance of these states can be due to the 
effects of the Cooper pair breaking, \cite{Band} the breaking mechanism 
being either intrinsic or the result of some structural defects that 
exist even in the high quality single crystals of YBCO synthesized 
until now. For example, Lee and Readon \cite{Lee} have shown 
that strong inelastic scattering of electrons results in the 
Cooper pair breaking. However, the origin of high scattering rate of 
electrons is not known in high temperature superconductors (HTSC).

Investigations of thermal conductivity of HTSC revealed that the heat 
conduction in these materials is mainly due to phonons: quite rough 
estimations of the phonon and electron contributions give 2/3 and 1/3 of 
the total thermal conductivity, respectively. \cite{Uher1} Such situation 
is fortunate to study the electron-phonon interaction in HTSC by the 
thermal conductivity method. It is important that thermal phonons interact 
with electrons even at temperatures below the critical temperature $T_{c}$. 
At temperatures not very lower than $T_{c}$, the phonon scattering rate is 
determined essentially by the electron-phonon interaction. In it's turn, 
the phonon relaxation is the basic factor that determines the lattice 
thermal conductivity. The phonon scattering rate in electron-phonon 
processes depends upon the amplitude of electron-ion interaction, 
the mass of charge carriers, and the energy dependence of the electron 
density of states $N(E)$ within a layer of the order of $k_{B}T$ near 
the Fermi level. The later is particular important for superconductors 
at $T<T_{c}$.

In the framework of a model of heat conduction that we follow, the
finite density of the in-gap states must results in reduction of 
thermal conductivity in the superconducting state. In this connection, the
investigation of thermal conductivity of YBCO under the controlled
variation of the Cooper pair breaking rate is important.

For our opinion, for such investigations, the YBCO with partial substitution
of copper by zinc is convenient. It is well known that even a small
concentration of zinc decreases $T_{c}$ considerably and, in the same 
time, makes unessential changes in lattice parameters. \cite{Mark} 
Neutron and calorimetric measurements
revealed that the phonon density of states in the low energy domain ($<15$
-- 20~meV) \cite{Par} and the Debue temperature $T_{D}$ \cite{Naka} does 
not practically change under zinc doping. This agrees with that the zinc
valency in YBCO, its ionic radius, and atomic mass almost coincide with
corresponding parameters of Cu$^{2+}$. From this one can expect that
zinc impurity in YBCO lattice is not a point defect for phonons. On the
other hand, according to the data on electric conductivity and Hall effect, 
\cite{Chien} zinc is an effective center for electron scattering; and
the doping produces unessential changes in the carrier concentration.
Experimental data of the tunneling spectroscopy, \cite{Akim} nuclear
magnetic resonance \cite{Ishi} suggest the existence of quasi-particle
in-gap states in the YBCO:Zn at $T\ll T_{c}$. This result is naturally
explained under the assumption that zinc addition gives rise to the
pair-breaking in YBCO.

Thus, comparing data on the thermal conductivity for pure YBCO, in which
the intrinsic mechanism of pair breaking exists, with YBCO:Zn, in which 
copper ions are partially substituted by zinc, the valuable information 
on the electron density of states in the superconducting state can be 
obtained. In this work, we present experimental data on the
temperature and magnetic field dependencies of thermal conductivity for 
YBCO single crystals with partial substitution of copper by zinc and discuse 
the obtained results.

\section{Samples and experimental}

Single crystals of YBCO:Zn were grown by the CuO flux method in Pt crucible. 
\cite{Byk} The crystals were grown under flux cooling from 1150$^{\circ }$C
to 860$^{\circ }$C in air. After this the crucible was removed from the furnace,
the flux pour out, and the crystals were quenched. Typical dimensions of
crystals were 1 -- 3~mm in the $ab$-plane and 10 -- 40~$\mu$m in the $c$%
-axis direction. Measurements of a chemical composition of samples
using an electron microscope with X-ray microanalizator have shown that Pt
concentration in YBCO:Zn crystals was smaller than 0.1~at.\%. The single
crystals were annealed in flowing oxygen at 600$^{\circ }$C during 1~h
following by the cool down to 300$^{\circ }$C with rate 10$^{\circ }$C/h.
According to the data of X-ray structural measurements, \cite{Alex} such
annealing regime provides the oxygen content of 6.93 -- 6.97 in YBCO. We 
did not measure the oxygen content in our crystals of YBCO:Zn.
However, according to results of many measurements performed on ceramic
samples of YBCO:Zn, the oxygen stoichiometry is independent (within an 
experimental error) upon zinc concentration up to $x=0.1$ and 
is $6.97\pm 0.01$ for annealing procedure similar to that we used 
(see, e.g., Ref. \onlinecite{Liang}). From this we suppose that the 
oxygen content in the annealed crystals of YBCO:Zn is practically the 
same and near 7.0.

Using an optical microscope, we observed a well developed twinning
structure on the crystals of YBCO:Zn. For thermal conductivity
measurements we selected crystals without traces of the flux on their
surface and with narrow superconducting transition width.

\begin{figure}[t]
 \begin{center}
  \leavevmode
  \epsfxsize=0.9\columnwidth \epsfbox {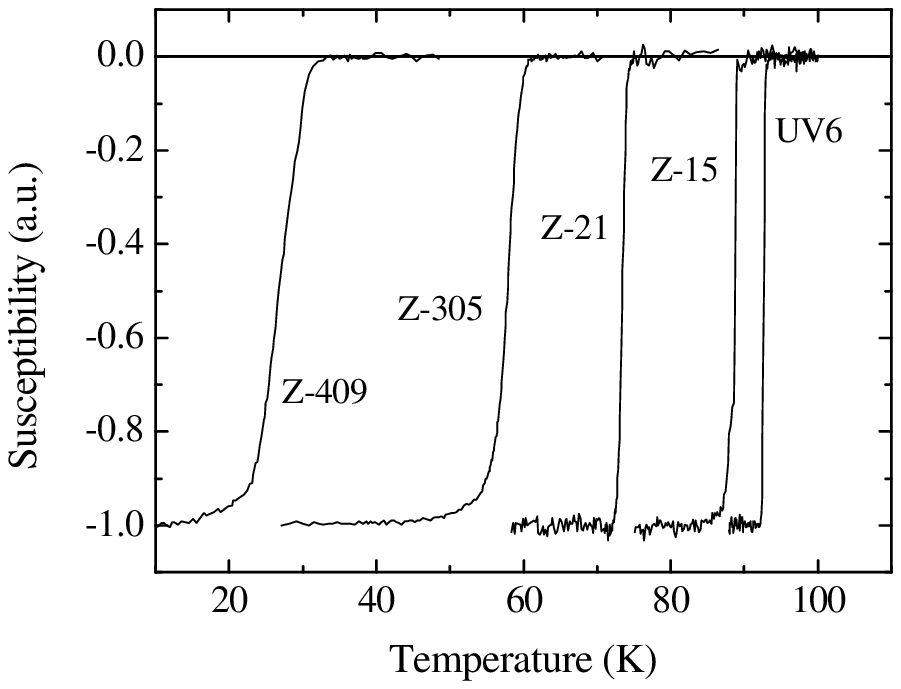}
  \caption{Temperature dependence of the magnetic susceptibility for YBa$_2$(Cu%
$_{1-x}$Zn$_x$)$_3$O$_{7-y}$ single crystals in ac magnetic field ${\bf H}%
\parallel {\bf c}$, $H=0.1$~Oe, $f=667$~Hz.}
  \label{fig1}
 \end{center}
\end{figure}

\begin{figure}[btp]
 \begin{center}
  \leavevmode
  \epsfxsize=0.9\columnwidth \epsfbox {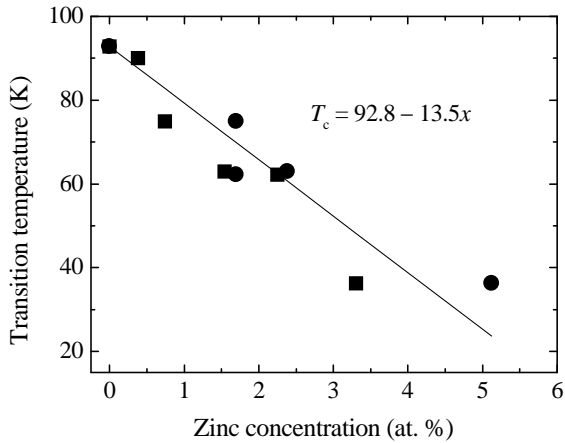}
  \caption{Dependence of the superconducting transition temperature of 
YBa$_2$(Cu$_{1-x}$Zn$_x$)$_3$O$_{7-y}$ single crystals upon the atomic 
concentration of zinc. Circles and squares represent the data of two 
different measurements of zinc concentration, a solid line is a linear 
approximation of the data.}
  \label{fig2}
 \end{center}
\end{figure}

The transition of samples into the superconducting state was detected using
the measurement of ac magnetic susceptibility (in magnetic field of
0.1~Oe at the frequency of 667~Hz) during the cool down from room
temperature. Fig. \ref{fig1} shows temperature dependence of magnetic 
susceptibility (a real part) for YBCO:Zn crystals, which we used 
for thermal conductivity measurements. It is seen that the width of
superconducting transition is narrow for YBCO samples with small zinc
concentration (0.4~K for pure YBCO and smaller than 1.4~K for samples
with $x<0.014$) and much broad for samples with $x>0.02$. In Table~1, 
the dimensions, critical temperature, transition width (determined
at the levels 0.1 and 0.9 of the maximum value of the magnetic
susceptibility), and zinc concentration are given for samples under
study. Zinc concentration in the samples with measured transition
temperatures was determined using a microprobe chemical analyzer. 
The obtained data are shown in Fig. \ref{fig2}. A linear approximation 
of the data gives $-13.5$~K/at.\% Zn for the rate of $T_{c}$ variation, 
which agrees well with the data of Ref. \onlinecite{Chien} for 
single crystals of YBCO:Zn.

\begin{table}[hbp]
\caption{Parameters of the YBa$_2$(Cu$_{1-x}$Zn$_x$)$_3$O$_{7-y}$ samples.}
\label{t1}
\begin{tabular}{lcccc}
\multicolumn{1}{c}{Sample} & \multicolumn{1}{c}{$x$} & 
\multicolumn{1}{c}{Dimensions,} & \multicolumn{1}{c}{$T_c$,} & 
\multicolumn{1}{c}{$\Delta T_c$,} \\ 
\multicolumn{1}{c}{} & \multicolumn{1}{c}{} & 
\multicolumn{1}{c}{mm} & \multicolumn{1}{c}{K} & 
\multicolumn{1}{c}{K} \\ 
\tableline UV-6 & 0.000 & $0.80\times 0.30\times 0.050$ & 92.8 & 0.4 \\ 
Z-15 & 0.003 & $0.90\times 0.30\times 0.040$ & 88.9 & 1.4 \\ 
Z-21 & 0.014 & $0.45\times 0.30\times 0.030$ & 74.0 & 1.2 \\ 
Z-305 & 0.025 & $0.45\times 0.32\times 0.014$ & 59.5 & 4.5 \\ 
Z-409 & 0.047 & $0.53\times 0.38\times 0.013$ & 30.0 & 7.0 \\ 
\end{tabular}
\end{table}

The thermal conductivity was measured using a temperature-wave method at
a frequency within the range from 0.2 to 2~Hz. The temperature wave in a
sample was generated with an electrical heater glued at the end of the 
sample. The temperature difference along the sample (less than 2\% of 
the mean temperature) was measured using a differential manganin-constantan
thermocouple. In a separate experiment, we had determined a magnetic field
dependence of the thermocouple sensitivity, and the values of thermal
conductivity in magnetic fields were corrected. The statistical scatter
of the data is about 0.1\% and the systematic experimental error is 50\%.
The later is mainly due to the error in measured value of a distance
between the thermocouple junctions, and in the heat current within 
the sample. The experimental technique will be published in more 
detail elsewhere.

\section{Results and discussion}

\subsection{Temperature dependencies of thermal conductivity}
\begin{figure}[tbp]
 \begin{center}
  \leavevmode
  \epsfxsize=0.9\columnwidth \epsfbox {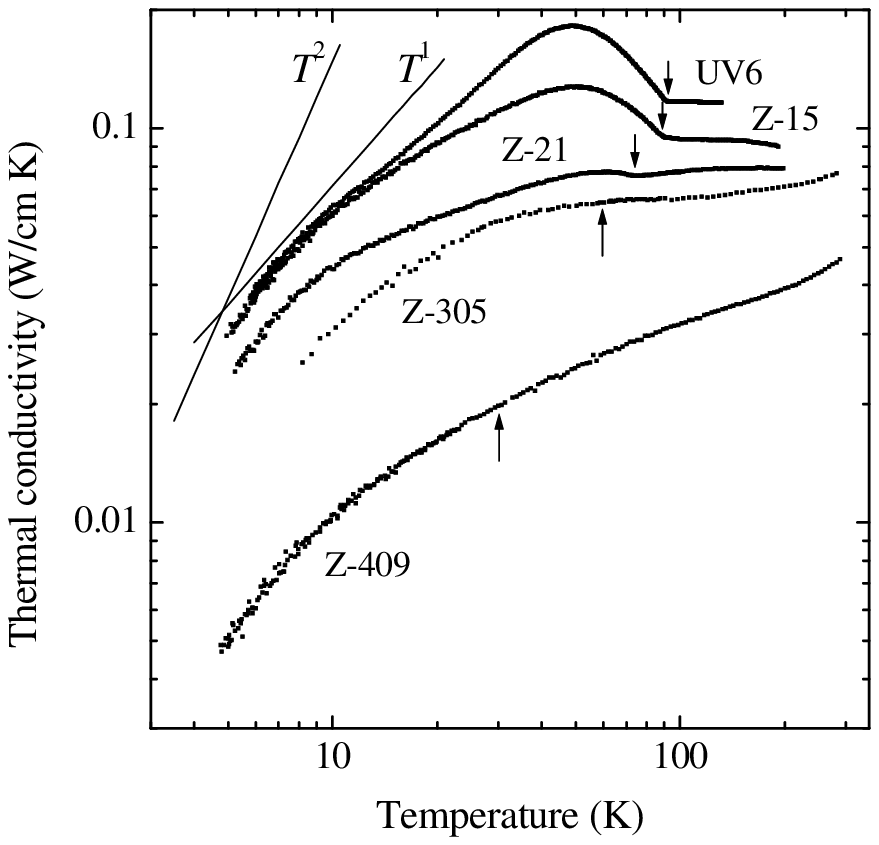}
  \caption{Temperature dependence of thermal conductivity for YBa$_2$(Cu$_{1-x}$%
Zn$_x$)$_3$O$_{7-y}$ single crystals along the $ab$-plane. Arrows show the
onset of superconducting transition.}
  \label{fig3}
 \end{center}
\end{figure}

The temperature dependencies of the in-plane thermal conductivity for 
YBCO:Zn single crystals are shown in Fig. \ref{fig3}. With increase Zn 
concentration we observe: (1) a decrease of thermal conductivity in 
the temperature domain studied, this effect being stronger at 
low temperatures; (2) a decrease in relative magnitude of 
the maximum in thermal conductivity $K_{\text{max}}/K(T_{c})$ 
down to its disappearance for samples with $x>0.025$.

In Ref. \onlinecite{Flor}, we studied the temperature and magnetic
field dependencies of thermal conductivity for YBCO single crystals.
It was shown that experimental data can be quite well account for on the
base of the theory, \cite{TW1} which is the modification of the Bardeen,
Rickaysen, and Tewordt (BRT) theory of thermal conductivity in
superconductors \cite{BRT} for the case of HTSC. We found that the phonon
heat transport is a dominant mechanism of thermal conductivity. In 
YBCO, the electron-phonon interaction determines the behavior of the phonon
thermal conductivity $K_{\text{ph}}$ to a considerable extent. With
temperature decrease below $T_{c}$, the electron thermal conductivity $K_{e}$
decreases rapidly, because electrons condensed into the Cooper pairs and
having a zero entropy do not participate in the heat transport, and the
concentration of normal electrons decreases with a temperature. At the
same time, the phonon relaxation due to the electron-phonon interaction
decreases dramatically, resulting in the increase of $K_{\text{ph}}$. The
rise of the total thermal conductivity with temperature decrease continues
until another mechanisms of the phonon relaxation (e.g., phonon scattering
by crystal lattice defects, by sample boundaries, etc.) become dominant. At
temperatures below the maximum, the thermal conductivity decreases mainly
due to the decrease of population of thermal phonons; and $K(T)$ is 
determined mainly by the particular frequency and
temperature dependence of the phonon relaxation rate.

Recently, Yu {\it et el.} \cite{Yu} proposed a quite different scenario 
of the increase of thermal conductivity in YBCO at $T<T_{c}$. They 
attributed this feature to the decrease of quasiparticle relaxation rate in
the superconducting state that gives rise to the dramatic increase in the
electron thermal conductivity. Among basic experimental facts, which form
the base of this model, \cite{Yu} there is an absent of peculiarities in
thermal conductivity of HTSC single crystals when the heat current is 
directed along the $c$-axis. This result was obtained by several groups. 
However, in the Ref. \onlinecite{Cao}, authors observed the increase of the
out-of-plane thermal conductivity in quite thin single crystals of yttrium
and thallium cuprates. Therefore, we consider that the model of Yu {\it et
el.} \cite{Yu} has rather unsteady experimental foundations, and we will
hold the phononic scenario. Until now, this (phonon) model explains all
experimental data on the thermal conductivity of HTSC, at list qualitatively.

Proceed from such ideas on the heat transport, we qualitatively analyze the
temperature dependence $K(T)$ of YBCO and its evolution upon copper
substitution by zinc.

As noted above, in pure YBCO, the electron thermal conductivity is about
30\% of the total at $T>T_{c}$. The observed considerable decrease of
thermal conductivity of Zn-doped YBCO by factor of 2 -- 4 for $x>0.02$ 
cannot be a result of decrease in $K_{e}$ only. Thus these experimental data
indicate that $K_{\text{ph}}$ in YBCO decreases considerably
with Zn doping. As the phonon scattering by lattice defects is not
the dominant mechanism of the thermal resistivity in the temperature range
considered, and the phonon-phonon scattering in YBCO:Zn does not change
essentially due to weak sensitivity of lattice parameters and lattice
dynamics to Zn impurities, then one can naturally assume that the observed
effects to be a result of the increase in the rate of phonon-electron
relaxation.

Previously, we have found \cite{Flor} that, for the pure YBCO, the using of
the following expression for the phonon-electron relaxation rate
\begin{equation}
\tau _{\text{ph-e}}^{-1}=A_{\text{ph-e}}xT,  \label{one}
\end{equation}
where the relaxation amplitude has the form \cite{Zai} 
\begin{equation}
A_{\text{ph-e}}\approx (k_{B}/\hbar )(m^{*})^2E_{\text{def}}^2/(2\pi D\hbar
^3 v_s),  \label{two}
\end{equation}
gives quite reasonable values for parameters involved: $m^{*}\approx 16m_{e}$
is an effective electron mass ($m_{e}$ is a free electron mass); $E_{%
\text{def}}\approx E_{F}\approx 0.3$~eV is a deformation potential, $D$ is
a specific density, and $v_{s}$ is a sound speed. From the expression
for $A_{\text{ph-e}}$ under the assumption that the deformation potential
changes slightly with Zn doping, it follows that the effective electron mass
increases, the increase being by factor of approximately two for $x=0.047$.

In the normal state, the electron thermal conductivity $K_{\text{en}}$ can
be estimated by using the formula: \cite{Zai} 
\begin{equation}
K_{\text{en}}=\sigma (L_{0}T),  \label{three}
\end{equation}
\begin{equation}
\sigma =\frac{ne^{2}\tau }{m_{\text{ab}}},  \label{four}
\end{equation}
where $\sigma $ is an electrical conductivity, $n$ is a carrier
concentration, $e$ is an electron charge, $m_{\text{ab}}$ is an effective
electron mass in the $ab$-plane, $\tau $ is an electron relaxation time,
and $L_{0}=(k_{B}/e)^{2}\pi ^{2}/3$ is the Lorenz number. On the base of
these formulae one can expect the twofold reduction of the electrical and
thermal conductivities of YBCO:Zn with $x=0.047$ (in comparison with pure
sample) as a result of increase of the effective electron mass, provided
that this increase is totally due to $m_{\text{ab}}$. At $T>T_{c}$ the value
of electron thermal conductivity is about 0.04~W/(cm~K). According to
the above presented analysis, the value of $K_{e}$ must be smaller than 0.02
W/(cm~K) and thus be less than 60\% of measured thermal conductivity at
100~K. This magnitude is overestimated because zinc addition causes the
decrease of the electron relaxation time.

The increase of effective electron mass should results in corresponding 
increase of a slope in the temperature dependence of electric 
conductivity (see Eq. \ref{four}). The authors of Ref. \onlinecite{Chien}, 
which deals with measurements of in-plane $\rho (T)$ for
single crystals of YBCO:Zn, have found that the slope ${\rm d}\rho /{\rm d}T$
increases by the factor of 1.7 as $x$ increases from 0 to 0.035. This agrees
with the increase of the mass $m^{*}$ inferred from the value of the lattice
thermal conductivity, which is estimated by using Eq. \ref{two}.

Let consider experimental data for low-temperature domain. In all
samples of YBCO:Zn at temperatures below about 15~K, the dependencies $K(T)$
are functionally the same: they follow the $T^1$ dependence and change it to 
$T^2$ law at lowest temperatures.

Usually, in dielectric crystals at $T\ll T_{D}$, the phonon thermal
conductivity is restricted by the sample dimensions due to phonon scattering
at sample boundaries; and many other mechanisms of phonon relaxation, such
as, for example, point defects, three phonon interactions, are ineffective. 
\cite{Ber} In the boundary scattering regime, $K(T)$ should vary as $T^3$.
But this does not agree with experimental data for single crystals of YBa$_2$%
Cu$_3$O$_{7-y}$ \cite{Grae} and Bi$_2$Sr$_2$CaCu$_2$O$_8$, \cite{Zhu} in
which the quadratic dependence of $K(T)$ is observed at low and ultralow
temperatures. In addition, the magnitude of thermal conductivity is much
less than theoretical one. As suggested in Refs. \onlinecite{Uher2,Peacor1},
the mechanism, which gives rise to the $T^2$ dependence, is the phonon
scattering by normal electrons. The existence of normal electrons at $T\ll
T_{c}$ has evidently no connection with thermally excited quasiparticles
above the gap, and is likely a consequence of finite density of states
within the gap.

Let suppose that it is the phonon-electron interaction determines the
low-temperature thermal conductivity of YBCO:Zn. As zinc concentration
increases, the thermal conductivity of YBCO:Zn decreases slightly at $T<15$%
~K (up to $x=0.003$, $T_{c}=88.9$~K), and then decreases much rapidly in
comparison with the behavior at high temperatures. For example, at 6~K the
thermal conductivity of the sample with $x=0.047$ ($T_{c}=30.0$~K) is 6.7
times smaller than the $K(6\text{ K})$ for pure YBCO, whereas at 100~K this
ratio is about 3.7. This behavior of the low-temperature thermal
conductivity indicates that, in the YBCO:Zn at $T\ll T_{c}$, the density of
states within the layer of $k_{B}T$ thickness at the Fermi level increases
with zinc concentration.

We clarify how the change in the quasiparticle density of states $N(E)$
influences the low-temperature thermal conductivity. According to the theory
of superconductors with paramagnetic impurities, \cite{Abri,Skal} the pick 
in the $N(E)$ becomes broader with increase of dopant concentration and 
there is a finite density of states near the Fermi level at some 
concentration. For low impurity concentration and at low temperatures, the 
energy of
thermal phonons is not enough high to excite quasiparticles. In this
situation the phonon-electron scattering rate does not increase and the
phonon thermal conductivity is practically the same as for pure
superconductor. It is this we observe for the sample with $x=0.003$. For
higher impurity concentration, when $N(E)$ is finite at any energy, the
phonon-electron relaxation channel works and this results in the decrease of
phonon thermal conductivity even at very low temperatures. The essential
decrease of the thermal conductivity of YBCO:Zn samples with relatively high
zinc content agrees with this picture. In this case, the allowable increase
of the electron thermal conductivity does not realized because defects even
of atomic scale strongly reduce $K_{e}$ in metals, but the phonon
thermal conductivity is weakly sensitive to points defects at $T\ll T_{D}$. 
\cite{Ber}

In a number of works dealt with experimental investigation of the point
contact spectra, \cite{Akim} specific heat, \cite{Loram} NMR and NQR on
copper and yttrium nuclei, \cite{Ishi,Allo} there are conclusions about the
existence of the gapless superconductivity in YBCO:Zn at $x>0.05$ and about
that the zinc impurity is an effective center for Cooper pairs breaking. The
authors of Refs. \onlinecite{Loram,Allo} supposed that the pair breaking in
YBCO:Zn has magnetic nature of the Abrikosov-Gor'kov type. \cite{Abri} Note,
the data of specific heat measurements in YBCO:Zn \cite{Loram} indicate that
there are in-gap excitations with practically zero energy even in samples
with low zinc concentration ($x\approx 0.01$), for which the $T_{c}$ decreases
slightly. This result disagrees with the theory, \cite{Abri} because it
(theory) predicts the gapless superconductivity at relatively high
pair-breaking rate when the critical temperature is much smaller than the
initial ``pure'' value. On the other hand, in recently published theoretical
work \cite{Band} dealt with pair-breaking effects in HTSC, the authors found
such unusual behavior of the $N(E)$ in the superconducting state under
strong inelastic pair-breaking scattering with even time-reversal symmetry.

Let now consider the experimental thermal conductivity data for intermediate
temperatures but below $T_{c}$. With increase of zinc concentration the
maximum in the thermal conductivity decreases and disappears for samples
with $x>0.02$. According to the theoretical models, \cite{TW1,TW2} which
agree well with experimental thermal conductivity data for YBCO, \cite
{Flor,Peacor2,Cohn} the magnitude of thermal conductivity $K_{\text{max}}$
in the maximum is determined mainly by the rate of phonon-electron
relaxation and phonon scattering from lattice defects. The higher the rate
of the normal electrons freeze out with temperature decrease, the higher 
$K_{\text{max}}$. Evidently, if the $N(E)$ is nonzero within the gap then
this results in the decrease of the $K_{\text{ph}}$ and in some increase of
the $K_{e}$. By the reasons mentioned above the gain in the total thermal
conductivity from the increase of electronic component is apparently
much smaller than the decrease of phonon component, and, as a result, the
decrease of $K_{\text{max}}$ is observed with zinc content rise. Thus, for
intermediate temperatures, experimental data agree with the ideas about
thermal conductivity behavior in superconductors with the Cooper pairs
breaking.

In the literature, as we know, there are no theoretical works on the
investigation of pair-breaking effects, e.g., caused by paramagnetic
impurities, on the {\it phonon} thermal conductivity of superconductors. 
In Ref. \onlinecite{Amb}, the theory of electron thermal conductivity of
superconductors with paramagnetic impurities was developed. Our estimations
of the phonon thermal conductivity for the case of finite density of states
at the Fermi level show that the dependence of $K_{\text{ph}}(T)$ is quite
sensitive to the form of the function $N(E)$, which strongly varies over the
energy range of the order of $k_{B}T$. This means that the calculations of
the $K_{\text{ph}}(T)$ requires the detail knowledge of the electronic
spectrum of the HTSC.

\subsection{Thermal conductivity in magnetic field}
At $T<T_{c}$, the thermal conductivity of pure YBCO decreases in the
magnetic field. \cite{Flor,Peacor3} We interpret this as a result of the
additional phonon scattering by normal electrons within the vortex cores. 
\cite{Flor}
\begin{figure}[thp]
 \begin{center}
  \leavevmode
  \epsfxsize=0.9\columnwidth \epsfbox {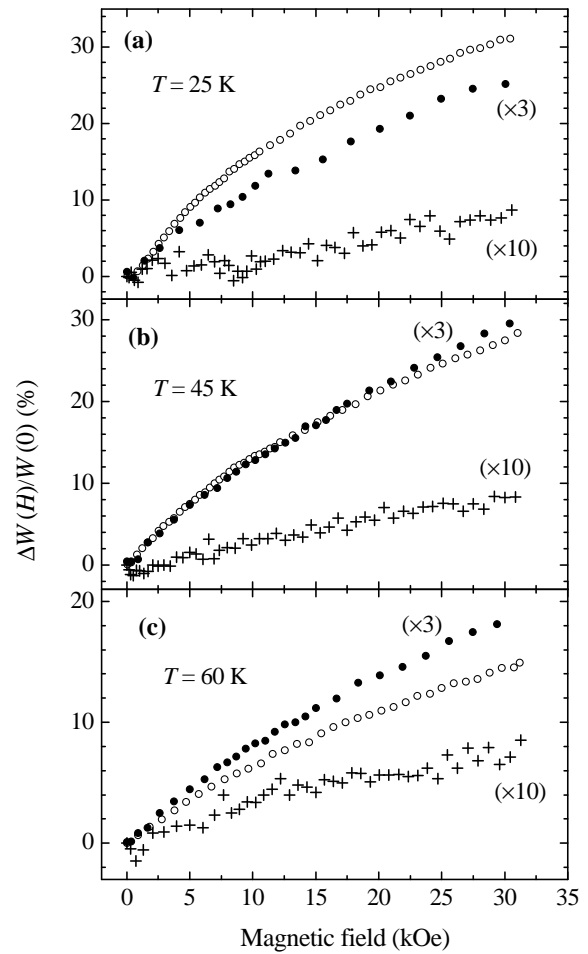}
  \caption{Excess in-plane thermal resistivity vs magnetic field 
${\bf H}\parallel {\bf c}$ for three YBa$_2$(Cu$%
_{1-x}$Zn$_x$)$_3$O$_{7-y}$ crystals at $T=25$~K (a), 45~K (b), and
60~K (c). $\circ $ -- sample UV-6, $x=0.0$; $\bullet$ -- sample Z-15, 
$x=0.003$; $+$ -- sample Z-21, $x=0.014$. The $\Delta W(H)/W(0)$ is
multiply by factor of 3 and 10 for samples Z-15 and Z-21, respectively.}
  \label{fig4}
 \end{center}
\end{figure}

In accord with the supposition about the increase of $N(E)$ at the Fermi
level in YBCO:Zn in the superconducting state, one can expect that the
increase of the concentration of normal component in the magnetic field will
take place at the background of already existing density of states, and,
thus, the higher doping level the weaker magnetic field effect. In Fig. \ref{fig4},
the field dependencies of the excess thermal resistivity $\Delta
W(H)/W(0)=(K(H)^{-1}-K(0)^{-1})K(0)$ are shown for three YBCO:Zn
samples with zinc concentration 0.0, 0.003, and 0.014 at temperatures 
25~K, 45~K, and 60~K for magnetic field orientation parallel to the $c$-axis 
of a crystal. The magneto-thermal resistivity deceases with increase of zinc
concentration: the effect smaller by factor of 2 -- 4 and of 20 -- 40 for $%
x=0.003$ and 0.014, respectively. The suppression of magnetic field effect
is larger at low temperatures. The obtained data generally agree with our
expectations. Note, that the reduction of the magnetic field effect with
zinc concentration correlates with the disappearance of the maximum in $%
K(T)$, because the phonon-electron interaction determines both of them.

\begin{figure}[btp]
 \begin{center}
  \leavevmode
  \epsfxsize=0.9\columnwidth \epsfbox {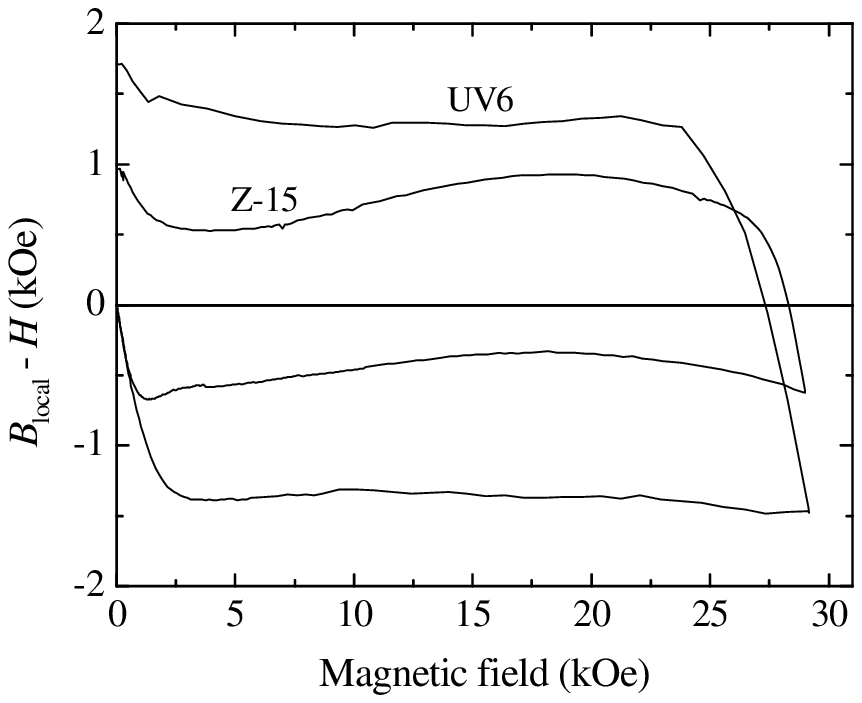}
  \caption{Magnetization curves for pure YBCO (UV6) and YBa$_2$(Cu$_{0.997}$Zn$%
_{0.003}$)$_3$O$_{7-y}$ (Z-15) in the magnetic field oriented parallel to
the $c$-axis at $T=25$~K.}
  \label{fig5}
 \end{center}
\end{figure}

At the same time, we note additional important (for our opinion) detail: in
the sample of YBCO:Zn with $x=0.003$, the lattice thermal conductivity 
is suppressed only slightly at $T<25$~K as seen from temperature dependence 
of $K(T)$. Therefore, the essential suppression of the magnetic field effect on
the thermal conductivity of YBCO:Zn must be determined by other factors. The
origins of such behavior are not clear now. It is possible that the rate of
phonon scattering by vortices depends upon the vortex pinning force.
Possibly, in pure YBCO, the substantial increase of the thermal resistivity
in magnetic field is connected with more strong pinning than in samples
doped with zinc. The later conclusion can be inferred from the absence of
the hysteresis phenomena in the YBCO:Zn samples. Moreover, our measurements
of the magnetization showed that at $T=25$~K the remnant magnetic moment $M_{%
\text{rem}}$ for the sample YBCO:Zn with $x=0.003$ is about two times
smaller than for pure YBCO (see Fig. \ref{fig5}) and, therefore, the pinning force is
smaller as well.

\section{Summary}
For the first time we have measured the temperature dependencies of 
thermal conductivity for YBa$_2$(Cu$_{1-x}$Zn$_x$)$_3$O$_{7-y}$ single 
crystals in the temperature range from 5 to 250~K. The copper 
substitution by zinc results in essential reduction of thermal 
conductivity and in the decrease of thermal conductivity pick 
down to its complete disappearance for samples with zinc 
concentration above 0.025. At temperatures below 15~K, dependencies 
of $K(T)$ for different samples are functionally the same.

Analysis of experimental data shows that in YBCO the effective electron
mass increases with Zn concentration rise, and the density of states 
within the gap (in the superconducting state) is nonzero even for 
smallest zinc concentrations. The later can not be account for in the 
frame of simple theory of superconductors with paramagnetic 
impurities. \cite{Abri,Skal} This suggests that some unusual mechanism 
of Cooper pair breaking may exist in YBCO:Zn crystals, the similar 
mechanism being in pure YBCO and resulting in the quadratic temperature 
dependence of thermal conductivity at low temperatures.

Introduction of zinc into YBCO causes the essential suppression of the
magnetic field effect on the thermal conductivity in the $ab$-plane in the
case when the field is directed perpendicular to the heat current. In 
contrast to the case of pure YBCO, no hysteresis phenomena are observed 
in the thermal resistivity. The comparative analysis of experimental data 
for different YBCO:Zn samples suggests that the pinning of the vortex 
line may results in the increase of the thermal resistivity in magnetic 
field.

\acknowledgements 
The authors thank S.~I.~Kucheiko and V.~V.~Frolov for chemical analysis of
samples and V.~V.~Florentiev for assistance in thermal conductivity
measurements. This work was supported by the Russian Scientific Council for
HTSC Problem (Grant No. 90389 of the State Program on the HTSC).


\begin{references}
\bibitem[*]{spct} This paper was published in Russian journal which is not 
easily accessible. The authors believe that the subject matter of the 
paper is steel interesting. Because of these, the authors have prepared 
the english version of the paper and placed it into the archive. 
Interpretation and discussion of data may be now out of date in some 
aspects, but the experimental data seem to be the issue of the day.

\bibitem{Gur}  M. Gurvitch, J. M. Valles, Jr., A. M. Cucolo, R. C. Dynes, J.
P. Garno, L. F. Schneemeyer, J. V. Waszczak, \prl {\bf 63}, 1008
(1989); J. M. Valles, Jr., R. C. Dynes, A. M. Cucolo, M. Gurvitch, L. F.
Schneemeyer, J. P. Garno, J. V. Waszczak, \prb {\bf 44}, 11986
(1991).

\bibitem{Coop}  S. L. Cooper, M. V. Klein, B. G. Pazol, J. P. Rice, D. M.
Ginsberg, \prb {\bf 37}, 5920 (1988).

\bibitem{Thom}  C. Thomsen, M. Cardona, B. Gegenheimer, R. Liu, A. Simon,
\prb {\bf 37}, 9860 (1988).

\bibitem{Boek}  M. Boekholt, M. Hoffmann, G. Guntherodt, Physica C {\bf 175}%
, 127 (1989).

\bibitem{Band}  C. Bandte, P. Hertel, J. Appel, \prb {\bf 45}, 8026 (1992).

\bibitem{Lee}  P. A. Lee and N. Read, \prl {\bf 58}, 2691 (1987).

\bibitem{Uher1}  C. Uher, J. Superconductivity {\bf 3}, 337 (1990).

\bibitem{Mark}  J. T. Markert {\it et al.}, in {\it Physical Properties of
High Temperature Superconductors}, ed. by D. M. Ginsberg (World Scientific,
Singapore, 1989), Vol. I, p.265.

\bibitem{Par}  P. P. Parshin, V. P. Glazkov, M. G. Zemlyanov, A. V. Irodova,
O. E. Parfenov, A. A. Chernishov, Superconductivity: phys., chem., tech. 
{\bf 5}(3), 451 (1992).

\bibitem{Naka}  Y. Nakazawa, J. Takeya, M. Ishikawa, Physica C {\bf 174},
155 (1991).

\bibitem{Chien}  T. R. Chien, Z. Z. Wang, N. P. Ong, \prl {\bf 67}, 
2088 (1991).

\bibitem{Akim}  A. I. Akimenko, G. Goll, I. K. Yanson, H. v. Lohneysen, R.
Ahrens, T. Wolf, H. Wuhl, Z. Phys. B {\bf 82}, 5 (1991).

\bibitem{Ishi}  K. Ishida, Y. Kitaoka, T. Yoshitomi, N. Ogata, T. Kamino, K.
Asayama, Physica C {\bf 179}, 29 (1991).

\bibitem{Byk}  A. B. Bykov {\it et al.}, J. Cryst. Grouth {\bf 91}, 302
(1988).

\bibitem{Alex}  I. V. Alexandrov, A. B. Bykov, I. P. Zibrov {\it et al.}
Pis'ma ZhETP {\bf 48}(8), 449 (1988).

\bibitem{Liang}  R. Liang, T. Nakamura, H. Kawaji, M. Itoh, T. Nakamura,
Physica C {\bf 170}, 307 (1990).

\bibitem{Flor}  V. V. Florentiev, A. V. Inyushkin, A. N. Taldenkov,
Superconductivity: phys., chem., tech. {\bf 3}(10), 2302 (1990).

\bibitem{TW1}  L. Tewordt, Th. Wolkhausen, Solid State Commun. {\bf 70}, 839
(1989).

\bibitem{BRT}  J. Bardeen, G. Rickaysen, L. Tewordt, Phys. Rev. {\bf 113},
982 (1959).

\bibitem{Yu}  R. C. Yu, M. B. Salamon, Jian Ping Lu, W. C. Lee, \prl
{\bf 69}, 1431 (1992).

\bibitem{Cao}  Shao-Chun Cao, Dong-Ming Zhang, Dian-Lin Zhang, H. M. Duan,
A. M. Hermann, \prb {\bf 44}, 12571 (1992).

\bibitem{Zai}  J. M. Zaiman, {\it Electrons and Phonons} (Clarendon Press,
Oxford, 1960), Ch.8, p.301.

\bibitem{Ber}  R. Berman, {\it Thermal Conduction in Solids} (Oxford, NY,
1976).

\bibitem{Grae}  J. E. Graebner, L. F. Schneemeyer, R. J. Cava, J. V.
Waszczak, E. A. Rietman, in {\it High-Temperature Superconductors}, ed. by
M. B. Brodsky {\it et al.}, 1987 MRS Fall Meeting Symposium Proceedings
(Materials Research Society, Pittsburgh, 1988), Vol.99, p.745-748.

\bibitem{Zhu}  Da-Ming Zhu, A. C. Anderson, E. D. Bukovski, D. M. Ginsberg,
\prb {\bf 40}, 841 (1989).

\bibitem{Uher2}  C. Uher, J. L. Cohn, J. Phys. C: Solid State Phys. {\bf 21}%
, L957 (1988).

\bibitem{Peacor1}  S. D. Peacor, C. Uher, \prb {\bf 39}, 11559 (1989).

\bibitem{Abri}  A. A. Abrikosov and L. P. Gor'kov, Zh. Eksp. Teor. Fiz. {\bf %
39}, 1781 (1960) [Sov. Phys. JETP {\bf 12}, 1243 (1961)].

\bibitem{Skal}  S. Skalski, O. Betbeder-Matibet, P. R. Weiss, Phys. Rev. 
{\bf 136}A, 1500 (1964).

\bibitem{Loram}  J. W. Loram, K. A. Mirza, P. F. Freeman, Physica C 
{\bf 171}, 243 (1990).

\bibitem{Allo}  H. Alloul, P. Mendels, H. Casalta, J. F. Marucco, J.
Arabski, \prl {\bf 67}, 3140 (1991).

\bibitem{TW2}  L. Tewordt, Th. Wolkhausen, Solid State Commun. {\bf 75}, 515
(1990).

\bibitem{Peacor2}  S. D. Peacor, R. A. Richardson, F. Nori, and C. Uher,
\prb {\bf 44}, 9508 (1991).

\bibitem{Cohn}  J. L. Cohn, S. A. Wolf, T. A. Vanderah, V. Selvamanickam, K.
Salama, Physica C {\bf 192}, 435 (1992).

\bibitem{Amb}  V. Ambegaokar, A. Griffin, Phys. Rev. {\bf 137}A, 1151 (1965).

\bibitem{Peacor3}  S. D. Peacor, J. L. Cohn, and C. Uher, \prb 
{\bf 43}, 8721 (1991).
\end{references}
\end{document}